\documentclass[authoryear,preprint,review,12pt]{elsarticle}

\usepackage{amssymb,amsmath}
\journal{Icarus}

\begin{document}

\begin{frontmatter}

  \title{Yarkovsky-driven impact risk analysis for asteroid (99942)
    Apophis}

  
  \author[jpl]{D. Farnocchia}
  \ead{Davide.Farnocchia@jpl.nasa.gov}
  \author[jpl]{S.~R. Chesley}
  \author[jpl]{P.~W. Chodas}
  \author[ifa]{M. Micheli}
  \author[ifa]{D.~J. Tholen}
  \author[unipi]{A. Milani}
  \author[ifa]{G.~T. Elliott}
  \author[spacedys]{F. Bernardi}

  \address[jpl]{Jet Propulsion Laboratory/California Institute of
    Technology, 4800 Oak Grove Drive, Pasadena, CA 91109, USA}
  \address[ifa]{Institute for Astronomy, University of Hawaii, 2680
    Woodlawn Drive, Honolulu, HI 96822, USA}
  \address[unipi]{Dipartimento di Matematica, Universit\`a di Pisa,
    Largo Pontecorvo 5, 56127 Pisa, Italy}
  \address[spacedys]{SpaceDyS, Via Mario Giuntini 63, 56023 Cascina,
    Pisa, Italy}

\begin{abstract}
  We assess the risk of an Earth impact for asteroid (99942) Apophis
  by means of a statistical analysis accounting for the uncertainty of
  both the orbital solution and the Yarkovsky effect.  We select those
  observations with either rigorous uncertainty information provided
  by the observer or a high established accuracy. For the Yarkovsky
  effect we perform a Monte Carlo simulation that fully accounts for
  the uncertainty in the physical characterization, especially for the
  unknown spin orientation. By mapping the uncertainty information
  onto the 2029 $b$-plane and identifying the keyholes corresponding
  to subsequent impacts we assess the impact risk for future
  encounters. In particular, we find an impact probability greater than
  10$^{-6}$ for an impact in 2068. We analyze the stability of the
  impact probability with respect to the assumptions on Apophis'
  physical characterization and consider the possible effect of the
  early 2013 radar apparition.
\end{abstract}

\begin{keyword}
  Asteroids, dynamics \sep Celestial mechanics \sep Near-Earth objects
  \sep Orbit determination
\end{keyword}

\end{frontmatter}

\section{Introduction}
Asteroid (99942) Apophis was discovered by R.~A. Tucker, D.~J. Tholen,
and F. Bernardi at Kitt Peak, Arizona on 2004 June 19 (Minor Planet
Supplement 109613). Because of its motion, the asteroid was
immediately recognized as an object of interest and therefore was
observed again on June 20. In the following days, issues related to
telescope scheduling, bad weather, and lunar interference made it
impossible to collect new observations. By the time the Moon was out
of the way, the solar elongation had become too small, thus making the
object even more difficult to observe. Thus, the asteroid was lost until
December 2004, when it was rediscovered by G.~J. Garradd at Siding
Springs, Australia \citep{recovery}. Thanks to the December 2004
observations, Apophis was recognized as a potentially hazardous
asteroid, with a peak impact probability (IP) of 2.7\% in April 2029
reported by Sentry\footnote{http://neo.jpl.nasa.gov/risk} and
NEODyS\footnote{http://newton.dm.unipi.it/neodys}. Also, the new
observations revealed astrometric errors in the discovery
observations, for which D.~J. Tholen remeasured the positions and
adjusted the time (Minor Planet Circulars 53585 and 54280). On 2004
December 27 the Spacewatch survey reported pre-discovery observations
\citep{precovery} from March 2004 that ruled out any impact
possibility for 2029.

In January 2005 Apophis was observed by radar from the Arecibo
observatory, obtaining three Doppler and two delay measurements. In
August 2005 and May 2006 two more Doppler observations were
obtained. These observations significantly reduced Apophis orbital
uncertainty and led to a 6 Earth radii nominal distance from the
geocenter in 2029 \citep{giorgini_apophis}.  Though the 2029 impact
was ruled out, other potential impacts were revealed in the following
decades. Moreover, the exceptionally small close approach distance
turns a well determined pre-2029 orbit to a poorly known post-2029
orbit for which scattering effects are dominant. Therefore, even small
perturbations prior to 2029 play an important role.

\citet{chesley_apophis06} shows that the Yarkovsky effect
\citep{bottke_yark} substantially affects post-2029 predictions, and
therefore has to be taken into account for Apophis impact
predictions. \citet{bancelin_apophis} also suggest that the Yarkovsky
effect is relevant, though they do not account for the corresponding
perturbation for their risk assessment. However, the available
physical characterization of Apophis is not sufficient to estimate the
Yarkovsky effect as was done in the case of asteroid (101955) 1999
RQ$_{36}$ \citep{milani_rq36,chesley_rq36}. In particular the lack of
information for spin orientation does not allow one to determine
whether the Apophis semimajor axis is drifting inwards or outwards. In
some cases, the orbital drift associated with the Yarkovsky effect can
be measured from the orbital fit to the observations
\citep{yarko_list}. Unluckily, this is not yet possible for Apophis,
and upcoming radar observations are unlikely to help in constraining
the Yarkovsky effect.  Another nongravitational perturbation that
could be important is solar radiation pressure. However,
\citet{apophis_radp} show that solar radiation pressure has a much
smaller effect on the Apophis trajectory than does the Yarkovsky
effect.

\section{Orbit determination}
\subsection{Dynamical model}
The scattering close approach in 2029 calls for a dynamical model that
is as accurate as possible.
We used an N-body model that includes the Newtonian attraction of the
Sun, the planets, the Moon, and Pluto. These accelerations are based
on JPL's DE424 planetary ephemerides \citep{de424}. We further added
the attraction of 25 perturbing asteroids, which are listed in
Table~\ref{t:perturbers}.
\begin{table}[t]
\begin{center}
\begin{tabular}{rlc}
\hline
IAU&Name&GM\\
No.    &                 &(km$^3$/s$^2$)\\
\hline
 1  &  Ceres           & 63.13$^a$\\
 4  &  Vesta           & 17.29$^b$\\
 2  &  Pallas          & 13.73$^c$\\
 10  &  Hygiea          &  5.78$^a$\\
511  &  Davida          &  2.26$^d$\\
704  &  Interamnia      &  2.19$^d$\\
 15  &  Eunomia         &  2.10$^d$\\
 3  &  Juno            &  1.82$^d$\\
 16  &  Psyche          &  1.81$^d$\\
 52  &  Europa          &  1.59$^d$\\
 88  &  Thisbe          &  1.02$^d$\\
 87  &  Sylvia          &  0.99$^d$\\
 6  &  Hebe            &  0.93$^d$\\
 65  &  Cybele          &  0.91$^d$\\
 7  &  Iris            &  0.86$^d$\\
 29  &  Amphitrite      &  0.86$^d$\\
532  &  Herculina       &  0.77$^d$\\
324  &  Bamberga        &  0.69$^d$\\ 
259  &  Aletheia        &  0.52$^d$\\
 11  &  Parthenope      &  0.39$^d$\\
 56  &  Melete          &  0.31$^d$\\
 14  &  Irene           &  0.19$^d$\\
419  &  Aurelia         &  0.12$^d$\\
 63  &  Ausonia         &  0.10$^d$\\
135  &  Hertha          &  0.08$^d$\\
\hline
\end{tabular}
\end{center}
\caption{Main belt asteroid perturbers and associated GM values. {\em Refs.}: $^a$\cite{bcm11},
  $^b$\cite{russell_dawn}, $^c$\cite{konopliv_etal_2011}, 
  $^d$\cite{carry}.}
\label{t:perturbers}
\end{table}

As was already pointed out by \citet{chesley_rq36} and
\citet{yarko_list}, the relativistic terms of the planets are
important when dealing with high precision orbit modeling. Therefore,
we used the Einstein-Infeld-Hoffman approximation \citep{moyer}, which
accounts for planetary relativistic effects.

\subsection{Astrometry}\label{s:astrometry}
We aimed at using only high quality astrometry. Therefore, we
selected observations from \citet{tholen_apophis}, for which rigorous
uncertainty information is provided by the observer.  In particular,
\citet{tholen_apophis} quantify the error due to different sources:
\begin{itemize}
\item astrometric fit error $\Delta\alpha_a$ and $\Delta\delta_a$;
\item centroiding error $\Delta\alpha_\chi$ and
  $\Delta\delta_\chi$;
\item tracking error $\Delta\alpha_t$ and $\Delta\delta_t$.
\end{itemize}
These observations were obtained by Mauna Kea and Kitt Peak
observatories and were reduced using the 2MASS star catalog
\citep{2MASS}. Though 2MASS has been recently used as a reference to
remove systematic errors present in other catalogs \citep{cbm10},
\citet{2MASS_bias} discuss the presence of biases in the 2MASS
catalog, and 0.02" systematic errors in declination are reported by
\citet{tholen_apophis}. Moreover, \citet{2MASS_PM} show how the
absence of proper motion in the 2MASS catalog is responsible for
regional biases detected in 2011 Pan-STARRS PS1 asteroid astrometry
\citep{known_server}. The bias in declination due to the absence of
proper motion is fairly close to a mean value 0.05". On the other hand
the right ascension bias is variable, getting as large as 0.14", and
shows a strong regional signature. Even so, 0.05" is a reasonable mean
for the right ascension bias too.  Therefore, to mitigate the absence
of proper motion, we relaxed the observational weights by adding a
time dependent uncertainty component related to proper motion biases
according to
\begin{equation}\label{eq:sigma_pm}
\sigma_{PM} = 0.05" \frac{yr - 2000}{2011 - 2000}\ ,
\end{equation}
where $yr$ is the year of the observation.

This time-dependent component of the bias takes into account the
uncorrected proper motion between the epoch of the observation and the
reference epoch of the 2MASS catalog (J2000). However, since the 2MASS
catalog has been obtained from images taken over a period of about 3
years, and not instantaneously at the reference epoch, even the
nominal J2000 positions in the catalog may be biased by an amount
directly related to how long before the year 2000 they were actually
observed by the survey. Therefore, we conservatively assumed an
additional $\sigma_B = 0.04$" error component that accounts for the
intrinsic bias of the catalog at the J2000 epoch.

By rolling up all the sources of error, we gave the
\citet{tholen_apophis} observations the following individual weights:
\begin{eqnarray}
  w_\alpha &=& 1/(\Delta\alpha_a^2 + \Delta\alpha_\chi^2 + 
    \Delta\alpha_t^2 + N\sigma_B^2 + N\sigma_{PM}^2)\label{e:w1}\\
  w_\delta &=& 1/(\Delta\delta_a^2 + \Delta\delta_\chi^2 + 
    \Delta\delta_t^2 + N\sigma_B^2 + N\sigma_{PM}^2)\label{e:w2}
\end{eqnarray}
where $N$ is the number of observations in the same night and accounts
for the fact that catalog biases of observations obtained in the same
region of the sky are expected to be highly correlated (since the same
stars were used in the astrometric solution). The results in
\citet{known_server} and \citet{2MASS_PM} show that the angular scale
of the correlation in the proper motion bias is larger than the
typical sky-plane motion of Apophis in one night. Therefore, the
temporal correlation of the biases likely extends to more than a
single night of observations, and $N$ should be increased to include
all observations within a certain angular distance of each other (up
to a few tens of degrees, and location-dependent). A better
statistical treatment would be to correct the observations by
subtracting the mentioned biases. However, this is beyond the scope of
the present paper and we plan to address this issue in the future.

In our analysis we also include a 2MASS-based position from the March
2004 precovery observations by the Spacewatch survey, remeasured with
the same techniques and tools described in \citet{tholen_apophis},
from a stack of five of the original exposures. Formal uncertainty
values are also available for this position, and are treated in the
same way.
%

Apophis was observed by the Arecibo radar telescope in 2005 and
2006. This radar apparition produced 2 delay and 5 Doppler
measurements for which rigorous uncertainty information is available
\citep {giorgini_apophis}.

The astrometry included so far covers the time interval from 2004
March 15 to 2008 January 9. However, Apophis has been further observed
through late 2012.
Therefore, we included observations from Magdalena Ridge (four
observations on 2011 March 14) and Pan-STARRS PS1 (four observations
on 2012 August 8 and four on 2012 December 29). These are the only two
observatories with a residual RMS for numbered asteroid observations
below 0.2" according to
AstDyS\footnote{http://hamilton.dm.unipi.it/astdys2}. Since the
Pan-STARRS PS1 observations are reduced using the 2MASS star catalog,
we relaxed weights consistent with \eqref{e:w1} and \eqref{e:w2}:
\begin{eqnarray}
  w_\alpha &=& 1/(0.30"^2 + N\sigma_B^2 + N\sigma_{PM}^2)\label{e:w1_b}\\
  w_\delta &=& 1/(0.30"^2 + N\sigma_B^2 + N\sigma_{PM}^2)\label{e:w2_b}
\end{eqnarray}
On the other hand, Magdalena Ridge observations are reduced using the
USNO-B1.0 catalog \citep{usno-b1.0}, which accounts for proper
motion. We therefore debiased according to \citet{cbm10} and weighted these
observations at 0.30". These twelve observations doubled the observed
arc with quality comparable to the \citet{tholen_apophis} astrometry.

\subsection{Orbital solution}
\begin{figure}[h]
\centerline{\includegraphics[width=10cm]{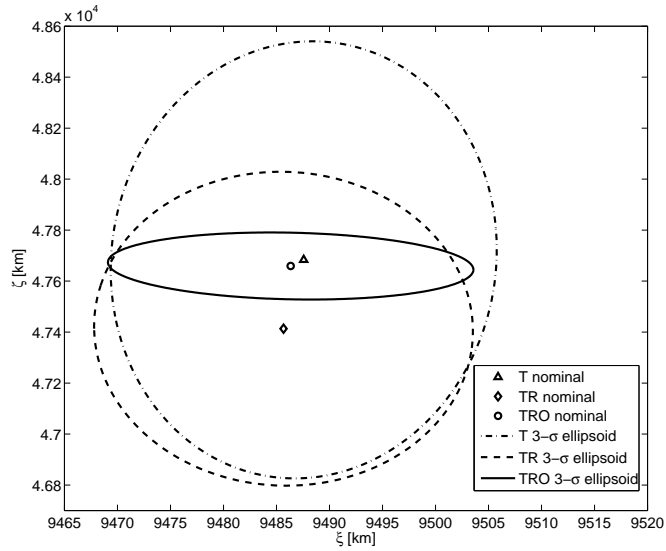}}
\caption{Nominal 2029 $b$-plane coordinates and 3-$\sigma$ uncertainty
  ellipses for different sets of observations: T is for only Tholen et
  al. observations, TR is for Tholen et al. plus radar observations,
  and TRO is for Tholen et al., radar, plus Pan-STARRS PS1 and
  Magdalena Ridge observations. Note that the axes do not use the same
  scale.}
\label{f:bplane}
\end{figure}

By using the observational data described in Sec.~\ref{s:astrometry}
we obtain the osculating orbital elements in Table~\ref{t:orbit}. We
then mapped the orbital elements and their uncertainties to the 2029
$b$-plane \citep{valsecchi}. Figure~\ref{f:bplane} shows the positions
on the $b$-plane corresponding to the orbital solutions: T is for only
Tholen et al. observations, TR is for Tholen et al.  plus radar
observations, and TRO is for Tholen et al., radar, plus Pan-STARRS PS1
and Magdalena Ridge observations.

\begin{table}[t]
\begin{center}
\begin{tabular}{lcc}
\hline
Epoch & 2006 Oct. 23.0 & TDB\\
Eccentricity & 0.1910634179(303) & \\
Perihelion distance & 0.7460469492(282) & au\\ 
Perihelion time & 2006 Jul. 7.8082161(101)& TDB\\ 
Longitude of Ascending Node & 204.4601395(387) & deg\\ 
Argument of perihelion & 126.3934237(392) & deg\\ 
Inclination & 3.33133146(361) & deg\\ 
\hline
\end{tabular}
\end{center}
\caption{Apophis' orbital parameters. Numbers in parentheses indicate
  the 1-$\sigma$ formal uncertainties 
  of the corresponding digits in the
  parameter value.}
\label{t:orbit}
\end{table}

Table~\ref{t:bplane} reports $b$-plane coordinates and uncertainties
for different choices of the observational dataset. (Note that the
$b$-plane is normal to the asymptotic velocity and the $(\xi,\zeta)$
coordinates mark the intersection of the asymptote with respect to the
geocenter.  Therefore, $b=\sqrt{\xi^2 + \zeta^2}$ does not correspond
to the closest approach, which is at about 39000 km.) We also included
the $b$-plane coordinates for the orbital solution obtained by using
all the available observations. In some cases there was an excess of
observations from a single observatory in a single night. In order to
reach the level of a preferred contribution of 5 observations per
night we relaxed the weights in these cases by a factor $\sqrt{N/5}$,
where $N$ is the number of observation in the night from the same
observatory.

\begin{table}[t]
\begin{center}
\begin{tabular}{lccccccc}
  Solution & N obs & $\xi$ & $\zeta$ & $\sigma_\xi$ &
  $\sigma_\zeta$ & $\Delta\xi$ & $\Delta\zeta$\\ 
  & &  km & km & km & km & km & km\\
  \hline
  T & 433 & 9487.57 & 47683.39 & 6.07 & 285.74 & -1.24 & -24.15\\
  TR & 440 & 9485.66 & 47413.94 & 5.95 & 205.20 & 0.67 & 245.30\\
  TRO & 452 & 9486.33 & 47659.24 & 5.75 & 43.84 & 0 & 0\\
  All & 1603 & 9482.98 & 47684.20 & 4.38 & 34.24 & 3.35 & -24.96\\
  \hline
\end{tabular}
\end{center}
\caption{2029 $b$-plane coordinates and uncertainties for different sets
  of observations: T is for only Tholen et al. observations, TR is for
  Tholen et al.
  plus radar observations, and TRO is for Tholen et al., radar, plus Pan-STARRS
  PS1 and  Magdalena Ridge observations, ALL is the solution obtained
  by using all the available observations. $\Delta\xi$ and $\Delta\zeta$
  are with respect to the TRO solution.}
\label{t:bplane}
\end{table}

It is worth noting that the T and TRO nominal solutions are
significantly shifted with respect to the TR prediction: the distances
in $\sigma$-space are 1.85 and 1.46, respectively. This difference may
point to the need for a better data treatment that consists in
correcting observations for catalog and proper motion biases, as
discussed in Sec.~\ref{s:astrometry}. It is also interesting to note
that the solution obtained with the full dataset is at 0.60 $\sigma$
from the TRO prediction and the improvement in the orbital uncertainty
is about 20\%. As \citet{polish_apophis} show that an appropriate
selection of the astrometric weighting scheme may be essential for
impact predictions, we prefer to use TRO as nominal solution, as we
have observations with a more reliable internal consistency and
uncertainty estimate.

\section{The Yarkovsky effect}\label{s:yarkovsky}
The Yarkovsky effect is a nongravitational perturbation that causes
asteroids to undergo a secular variation in semimajor axis, resulting
in a quadratic in time runoff in anomaly.

As described in \citet{yarko_list}, we modeled the Yarkovsky effect as
a purely transverse acceleration $a_t = A_2/r^2$, where $r$ is the
heliocentric distance and $A_2$ is a function of the asteroid's
physical quantities. The corresponding semimajor axis drift can be
computed as
\begin{equation}\label{e:dadt}
\frac{da}{dt} = \frac{2 A_2 (1 - e^2)}{n p^2}\ ,
\end{equation}
where $e$ is the eccentricity, $n$ is the mean motion, and $p$ is the
semilatus rectum.

\begin{figure}[h]
\includegraphics[width=7cm]{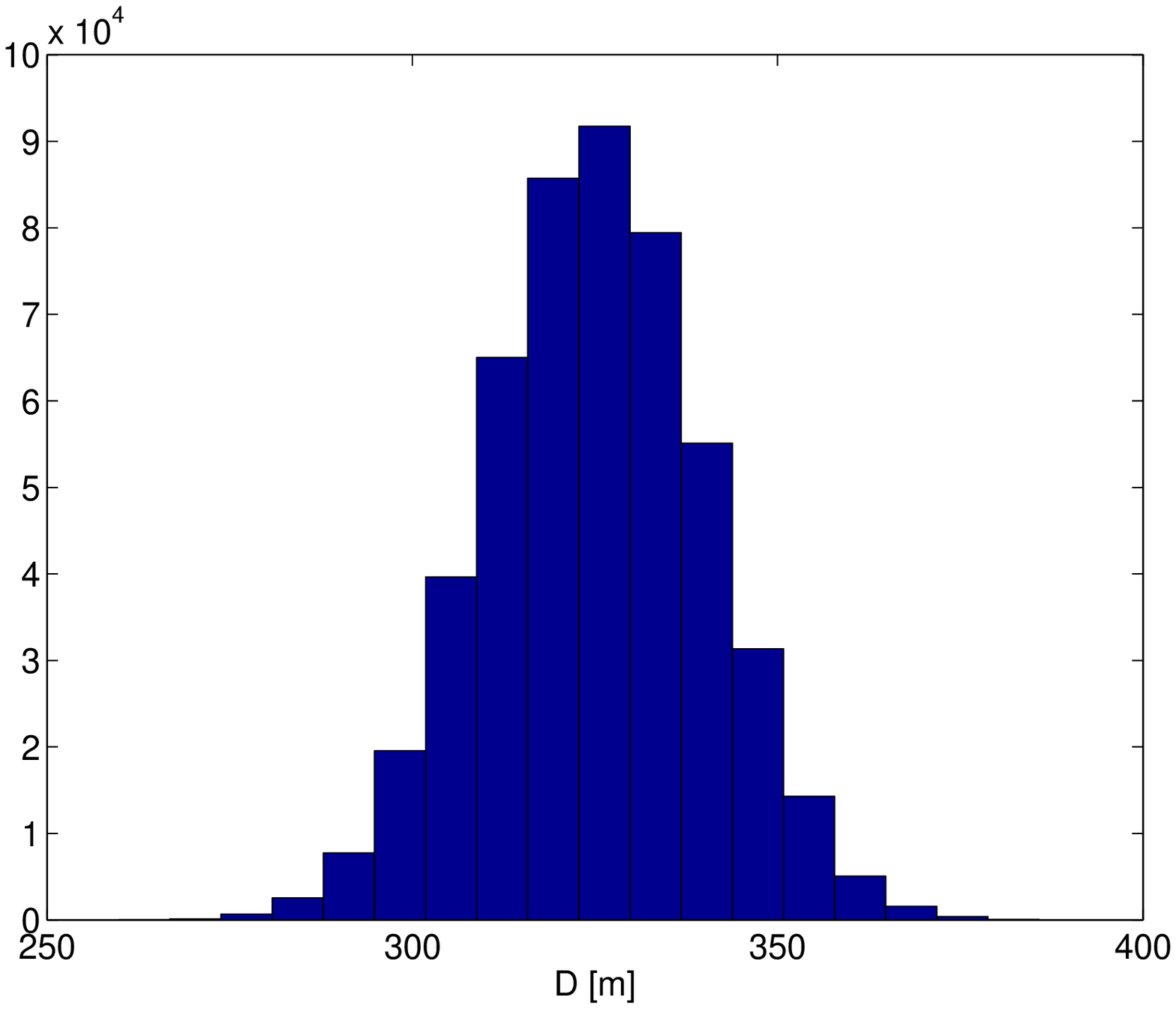}\includegraphics[width=7cm]{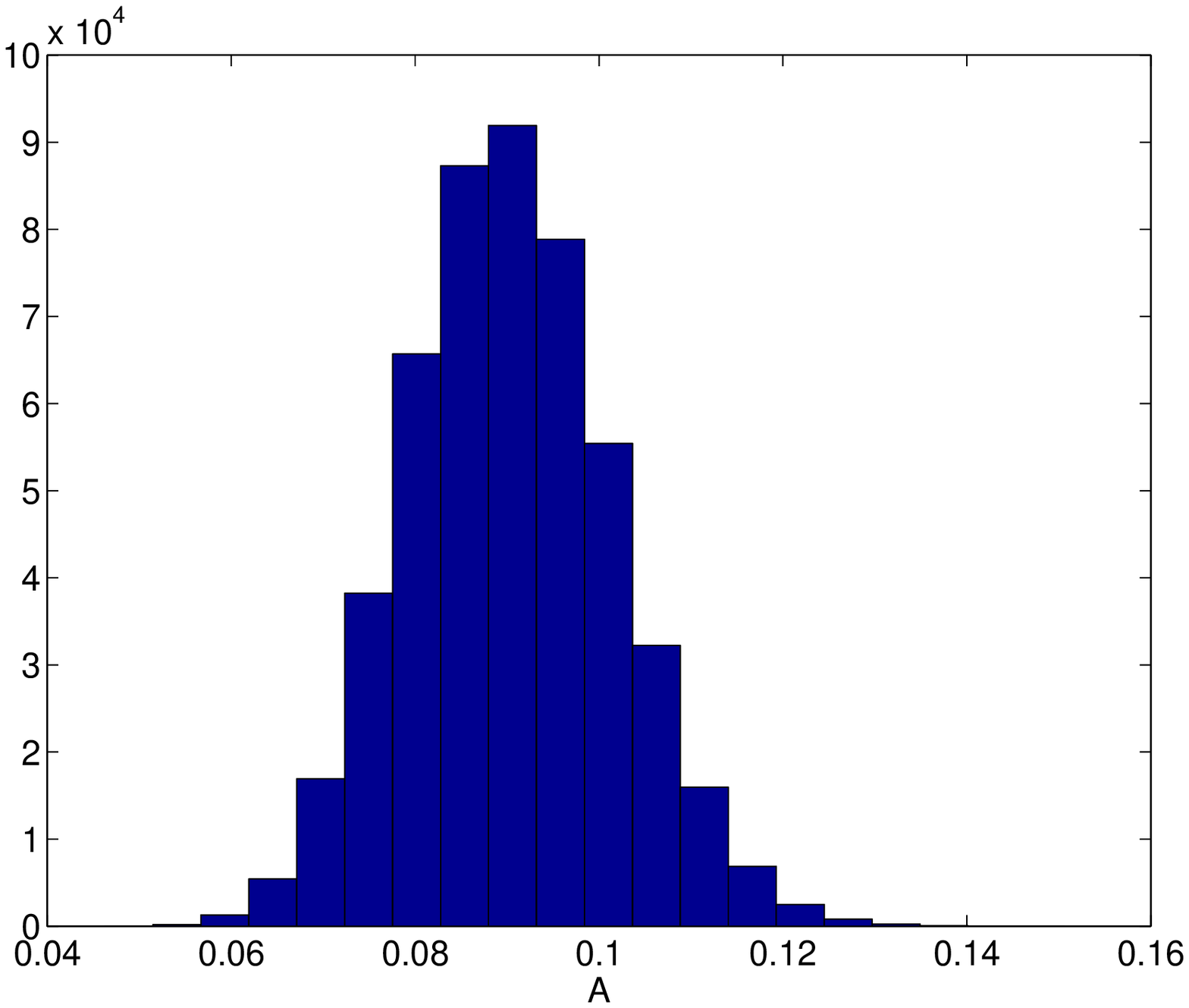}
\includegraphics[width=7cm]{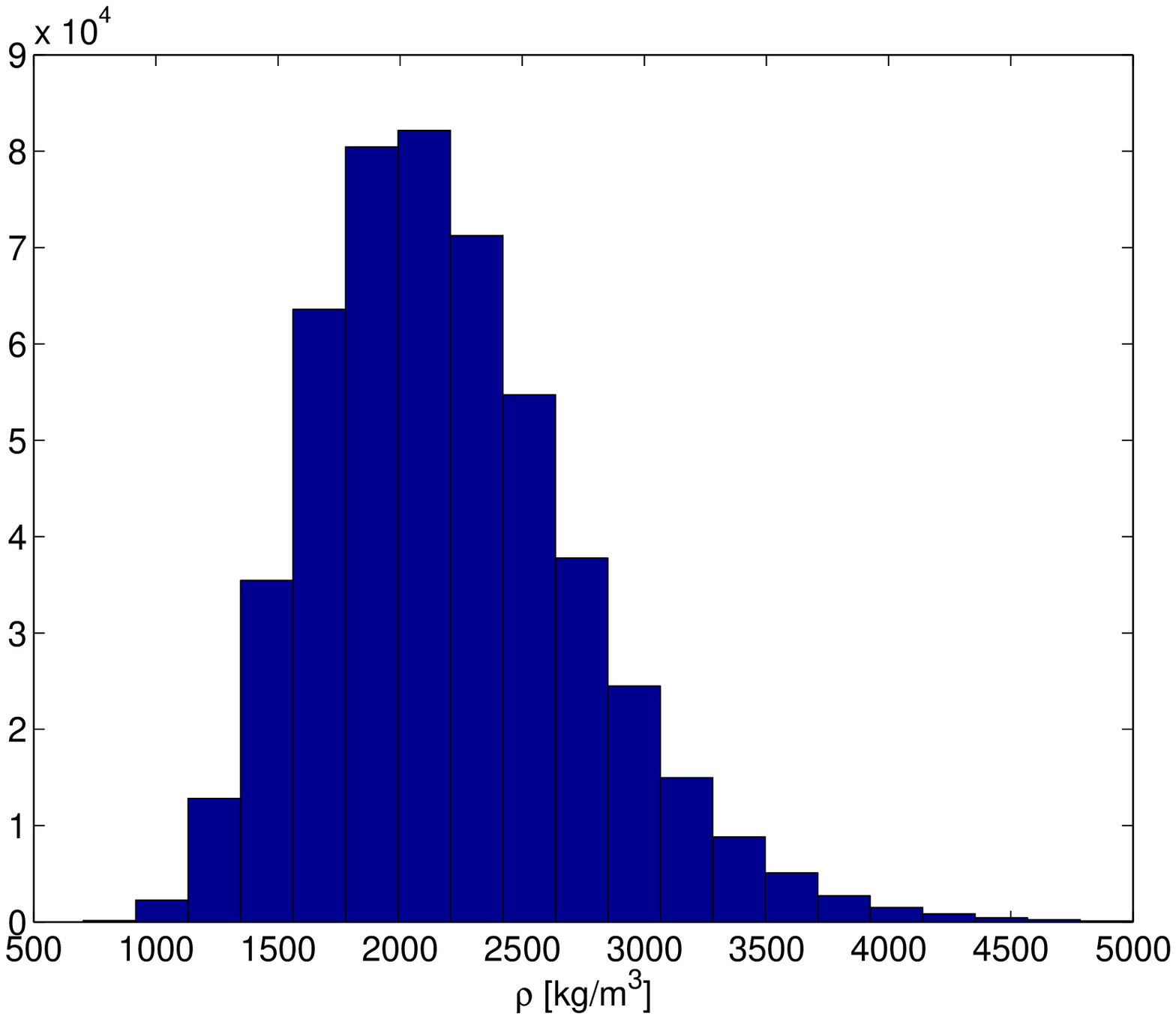}\includegraphics[width=7cm]{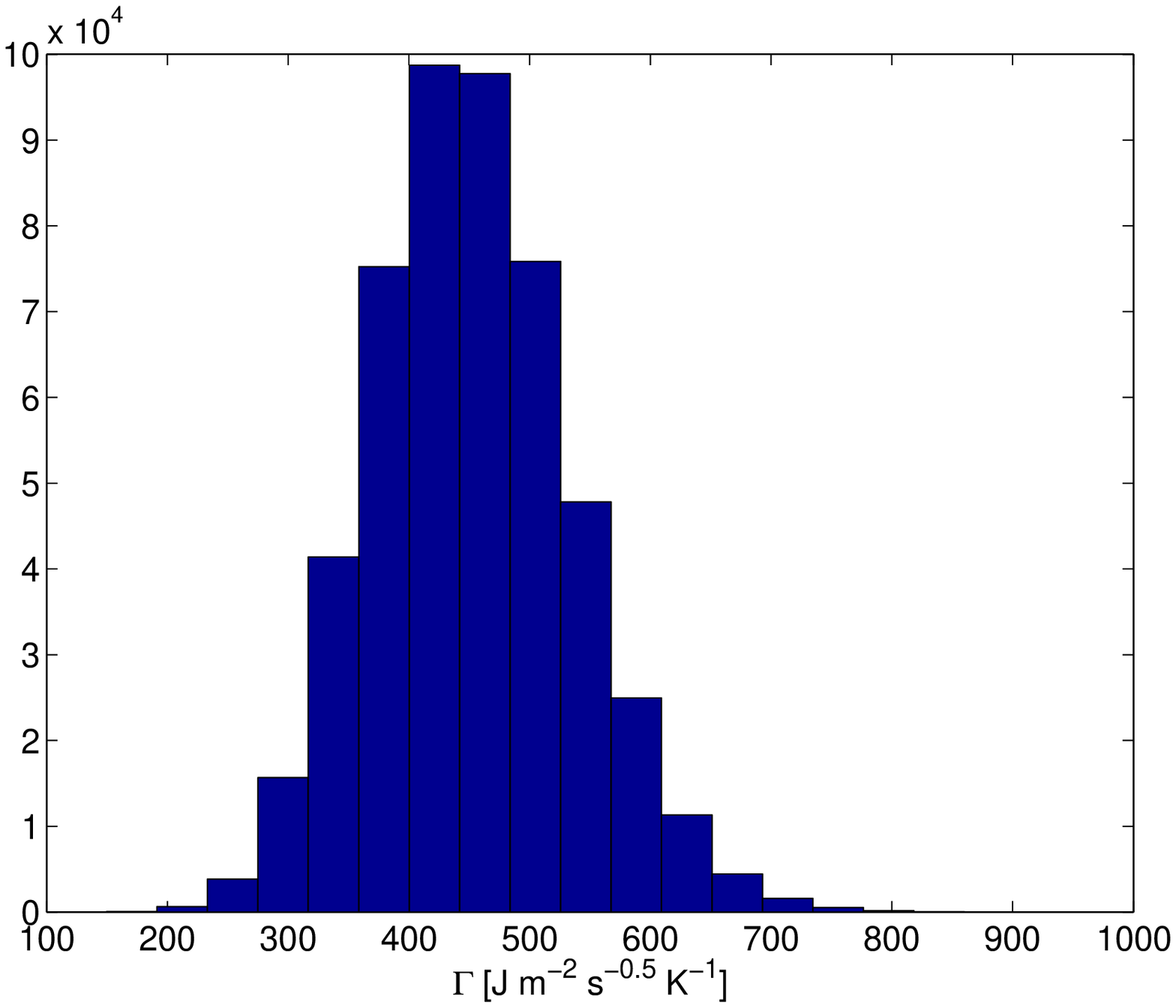}
\includegraphics[width=7cm]{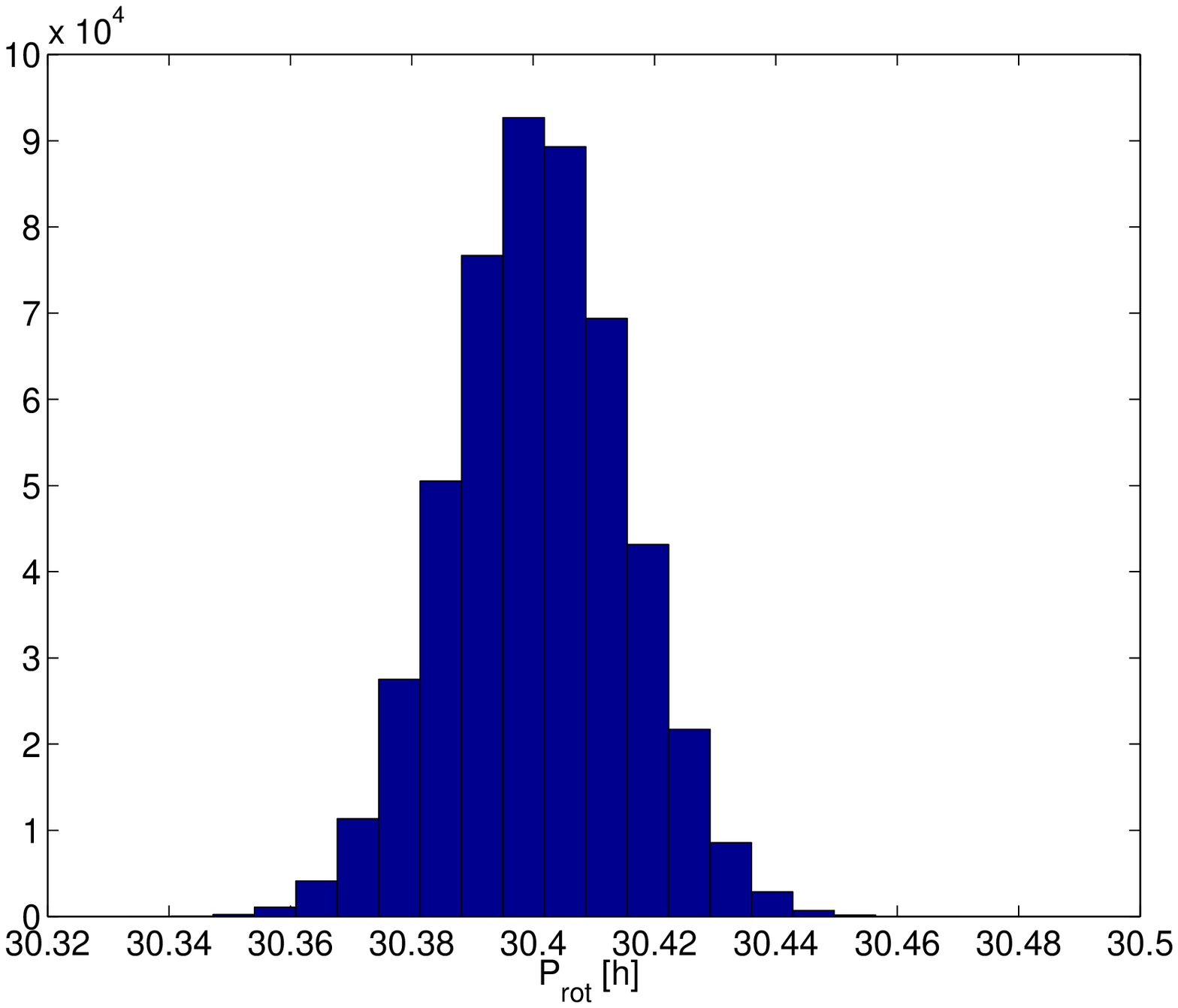}\includegraphics[width=7cm]{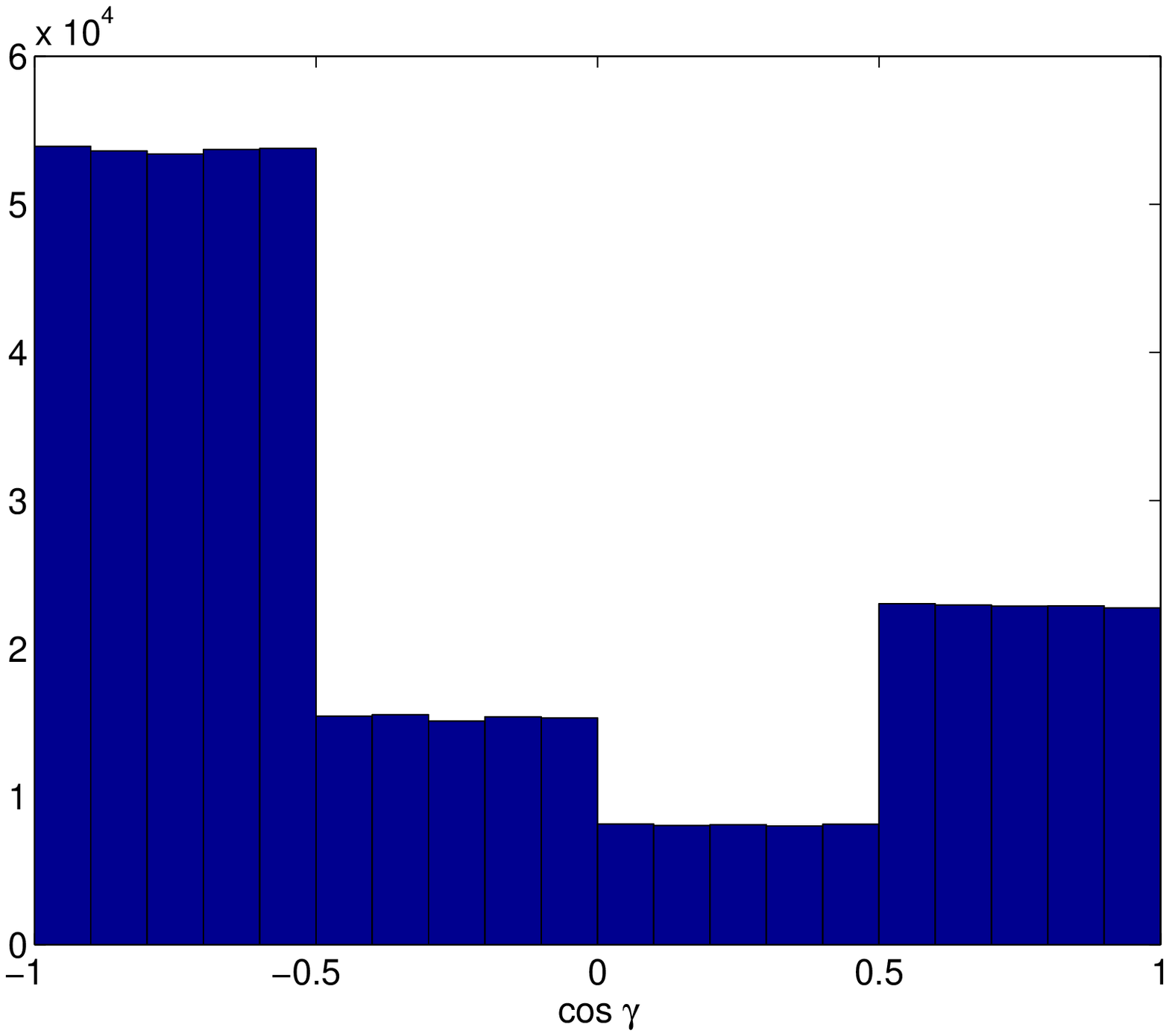}
\caption{Sampling of Apophis diameter $D$, Bond albedo $A$, density
  $\rho$, thermal inertia $\Gamma$, rotation period $P_{rot}$, and
  obliquity $\gamma$.}
\label{f:physical}
\end{figure}

As mentioned, $A_2$ can be neither computed from the available
physical characterization nor estimated from the orbital fit. In fact,
with current astrometry we get an uncertainty $\sigma_{A_2} = 70
\times 10^{-15}$ au/d$^2$ while the likely range of $A_2$ for an
object of this size is $\pm 25 \times 10^{-15}$ au/d$^2$. Therefore,
to compute the distribution of the semimajor axis we performed a Monte
Carlo simulation as was previously suggested by
\citet{chesley_apophis09}. Assumed physical parameters for the
simulation were obtained as follows:
\begin{itemize}
\item Recent observations from ESA's Herschel space
  observatory\footnote{http://www.esa.int/Our\_Activities/Space\_Science/Herschel}
  give a diameter $D = (325 \pm 15)$ m (T. Mueller, personal
  communication).  Thus, we used a normal distribution with mean 325 m
  and standard deviation 15 m (see top left panel of
  Fig.~\ref{f:physical}). Herschel observations also provide a
  geometric albedo $p_V = 0.23 \pm 0.02$. By assuming a slope
  parameter $G = 0.15\pm 0.05$, we converted $p_V$ to the Bond albedo
  $A =(0.29 + 0.684 G) p_V$ \citep{bowell}.  The top right panel of
  Fig.~\ref{f:physical} shows the distribution of $A$.
\item \citet{binzel} report a grain density between 3.4 and 3.6
  g/cm$^3$ and a total porosity between 4\% and 62\%. This leads to a
  density $\rho$ between 1.29 and 3.46 g/cm$^3$.  Therefore, we
  assumed a lognormal distribution with mean 2.20 g/cm$^3$ and
  variance 0.3 g$^2$/cm$^6$ (see middle left panel of
  Fig.~\ref{f:physical}). This distribution is consistent with the
  constraints of \citet{binzel}, as the probability of being smaller
  than 1.14 g/cm$^3$ or larger than 3.24 g/cm$^3$ is 2.03\% and
  2.47\%, respectively.
\item For Apophis the thermal inertia has not yet been measured.
  \citet{delbo_thermal} find a generic relationship between the
  thermal inertia and the diameter, according to which $\Gamma=d_0
  D^{-\psi}$, with $d_0 =( 300 \pm 45)$ J m$^{-2}$ s$^{-0.5}$
  K$^{-1}$, $D$ is in km, and $\psi = 0.36 \pm 0.09$. Accordingly, we
  used normal distributions for $\Gamma$ and $\psi$, and combined them
  with the already available diameter distribution. The middle right
  panel of Fig.~\ref{f:physical} shows the distribution of the thermal
  inertia.
\item Behrend's asteroids and comets rotation curves
  website\footnote{http://obswww.unige.ch/$\sim$behrend/page\_cou.html}
  reports a rotation period $P_{rot} = (30.4008 \pm 0.0144)$
  h\footnote{http://obswww.unige.ch/$\sim$behrend/r099942a.png},
  therefore we used the corresponding normal distribution (bottom left
  panel of Fig.~\ref{f:physical}).
\item The obliquity of the Apophis rotational pole is
  unknown. Assuming a uniform distribution for the obliquity is a poor
  approximation. In fact, according to \citet{laspina} retrograde and
  direct rotators are in a 2:1 ratio within the NEO population. As
  shown in the bottom right panel of Fig.~\ref{f:physical}, we used
  the obliquity distribution from \citet{yarko_list}, where the
  obliquity is inferred from a list of detections of the Yarkovsky
  effect.
\end{itemize}

\begin{figure}[h]
\includegraphics[width=7cm]{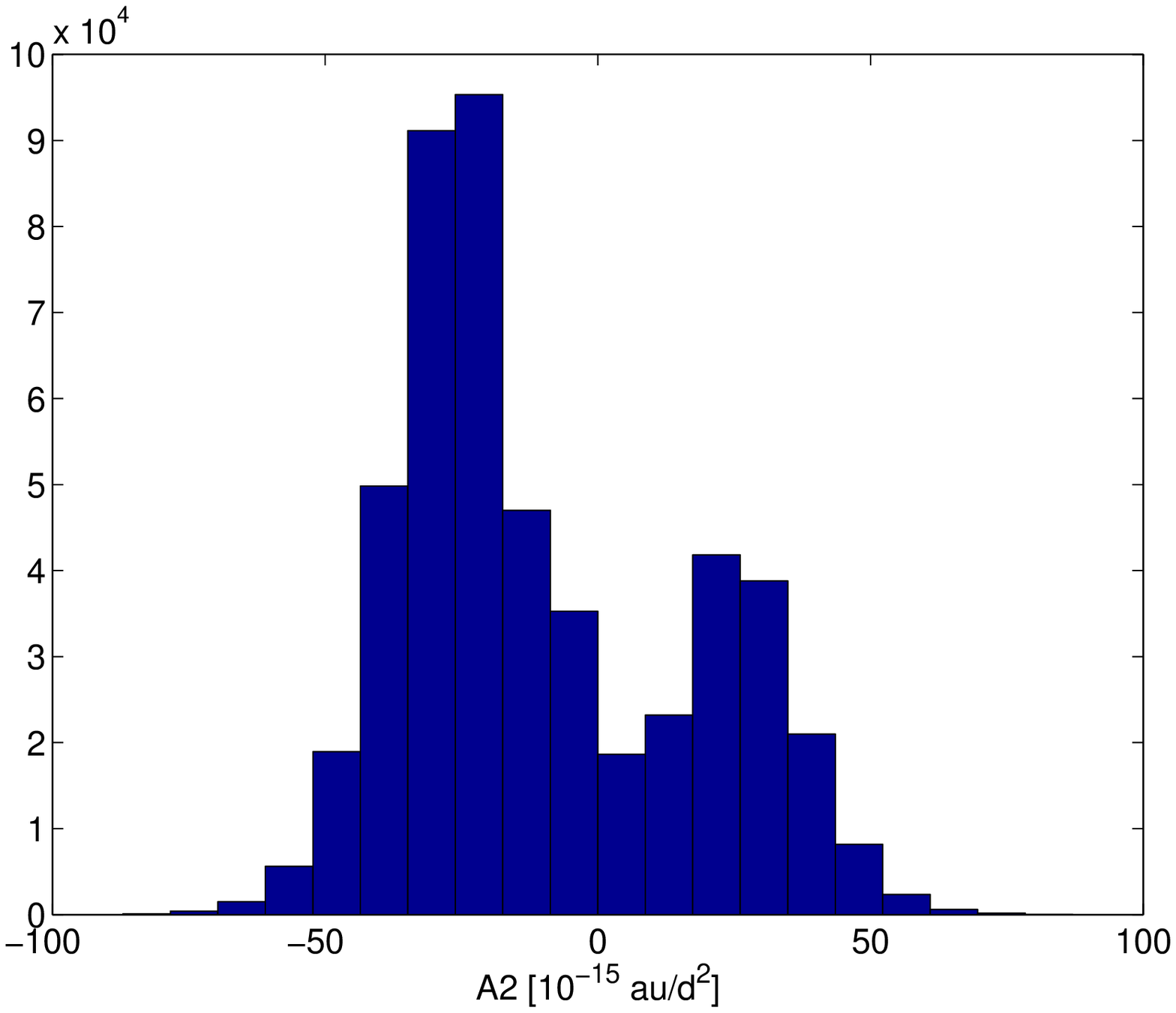}\includegraphics[width=7cm]{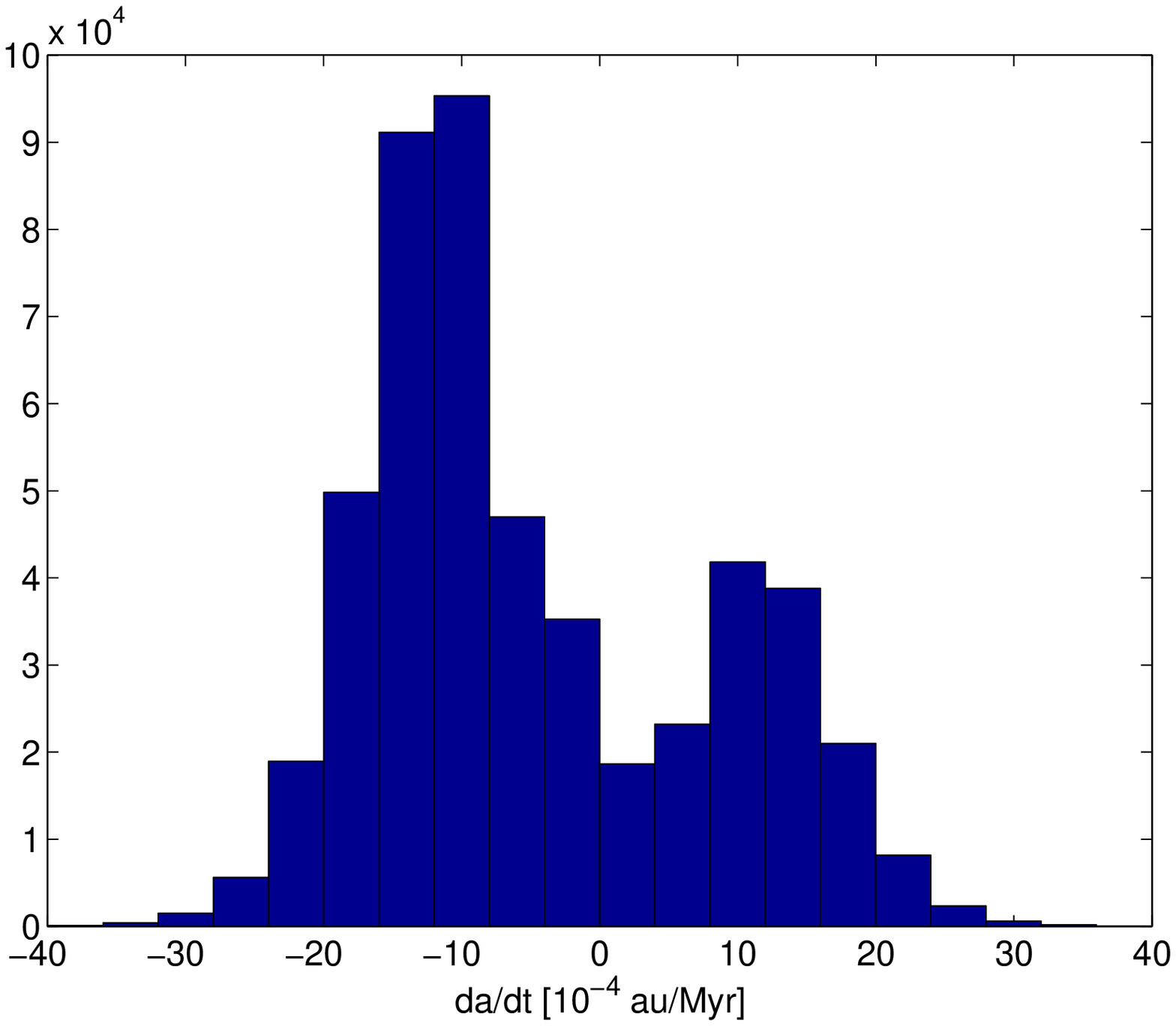}
\caption{Distribution of $A_2$ and $da/dt$ obtained from the assumed
  physical modeling.}
\label{f:yarko_par}
\end{figure}

For each point in the Monte Carlo sampling we computed the associated
$A_2$ according to \citet{yarko_list}, which is based on the linear
model of \citet{vok_2000}, and the orbital drift was then obtained by
\eqref{e:dadt}. Figure~\ref{f:yarko_par} shows the distribution of
$A_2$ and $da/dt$. The bimodality of the distributions corresponds to
the extreme obliquity peaks.

As previously done for the orbital solution uncertainty, we want to
see how the Yarkovsky effect maps to the 2029 $b$-plane.  To do so we
assumed an arbitrary reference value $(A_2)_{ref} = 10 \times
10^{-15}$ au/d$^2$, corresponding to $(da/dt)_{ref} = 4.59 \times
10^{-4}$ au/Myr, for which we obtained $\Delta\zeta_{ref} = 138.18$ km
and $\Delta\xi_{ref} = 0.45$ km. This confirms that the Yarkovsky
perturbation essentially affects only the $\zeta$ coordinate, which is
strongly related to the time and then to the anomaly
\citep{valsecchi}. To compute the value of $\Delta\zeta$ for a generic
point of the Monte Carlo sampling we scaled from the reference value,
after testing the validity of the linear approximation.

%

\section{Keyholes and risk assessment}
\begin{figure}[h]
\centerline{\includegraphics[width=14cm]{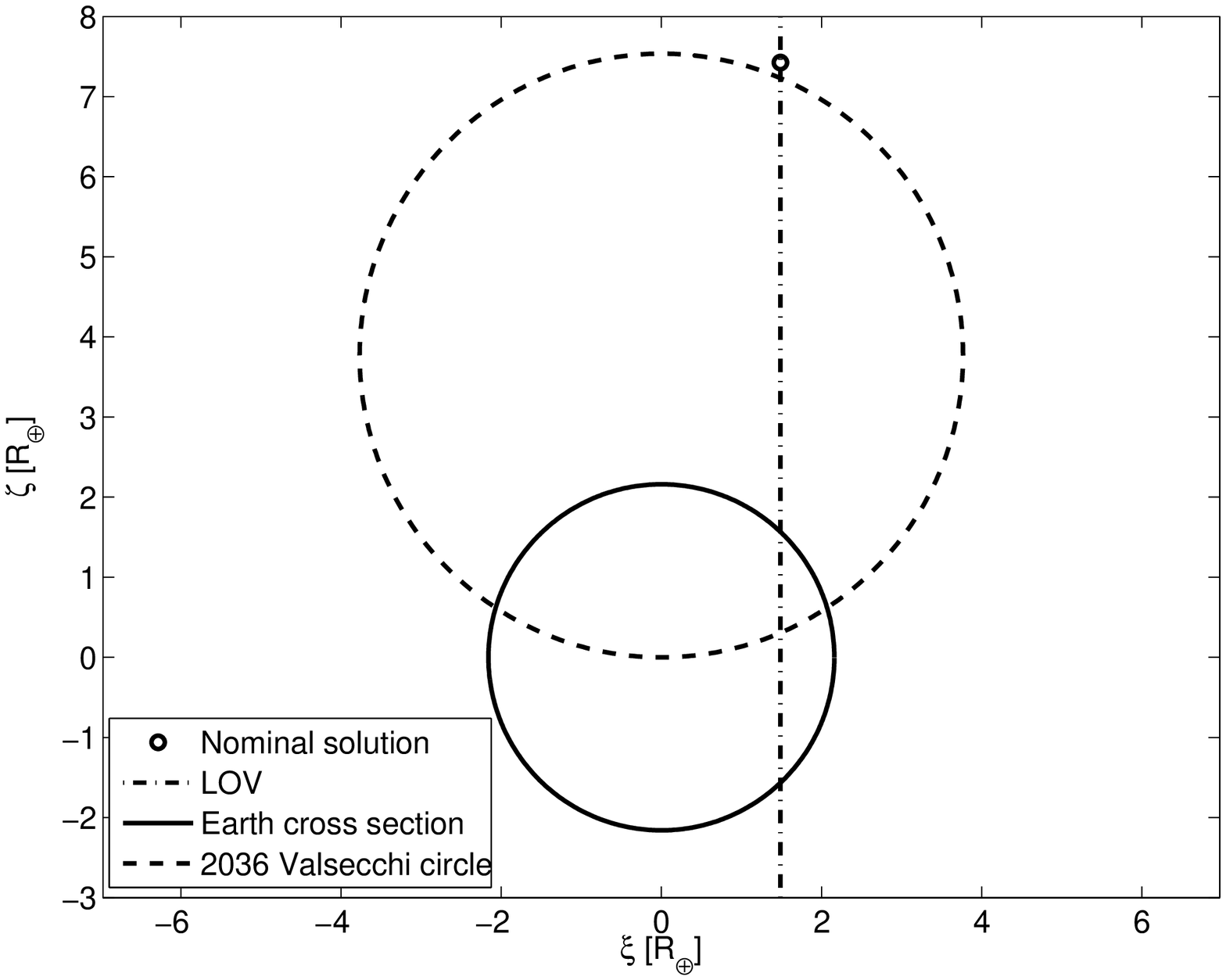}}
\caption{2036 resonant return, LOV, and keyhole on the 2029
  $b$-plane. The impact cross section of the Earth is 2.16
  $R_\oplus$.}
\label{f:resonant_return}
\end{figure}

Even though the 2029 close approach does not lead to an impact, we can
use the corresponding $b$-plane analysis to predict future
impacts. \citet{valsecchi} show that the $b$-plane coordinates leading
to a resonant impact lie on predictable
circles. Figure~\ref{f:resonant_return} shows the Valsecchi circle on
the 2029 $b$-plane for the 2036 resonant return, corresponding to a
7:6 resonance. The Line of Variation \citep[LOV,][]{lov1} is
representative of the orbital uncertainty region.  The intersection
between the orbital uncertainty region and a Valsecchi circle is a
keyhole \citep{keyholes}, leading to a future impact.

\begin{figure}[h]
\centerline{\includegraphics[width=12cm]{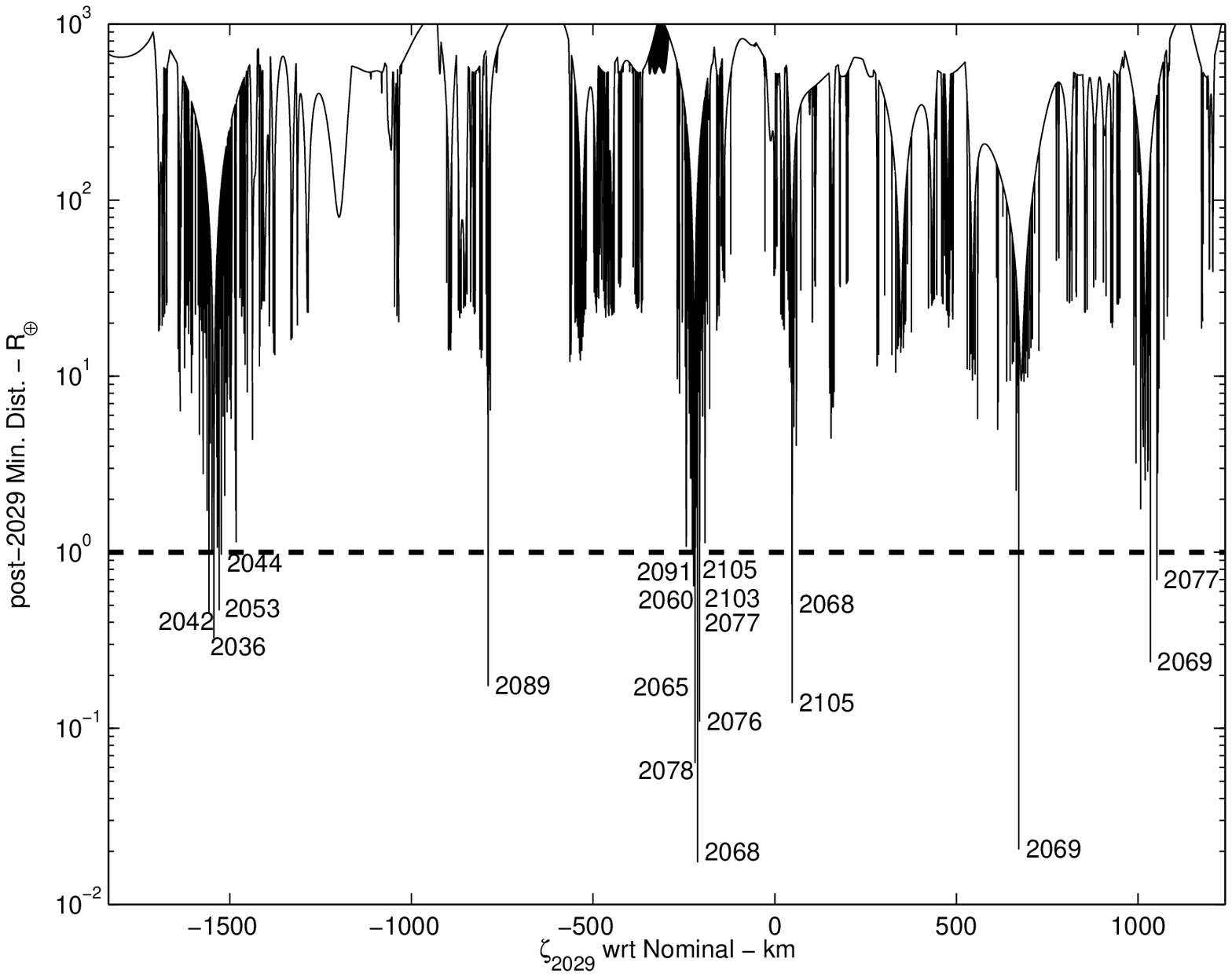}}
\centerline{\includegraphics[width=12cm]{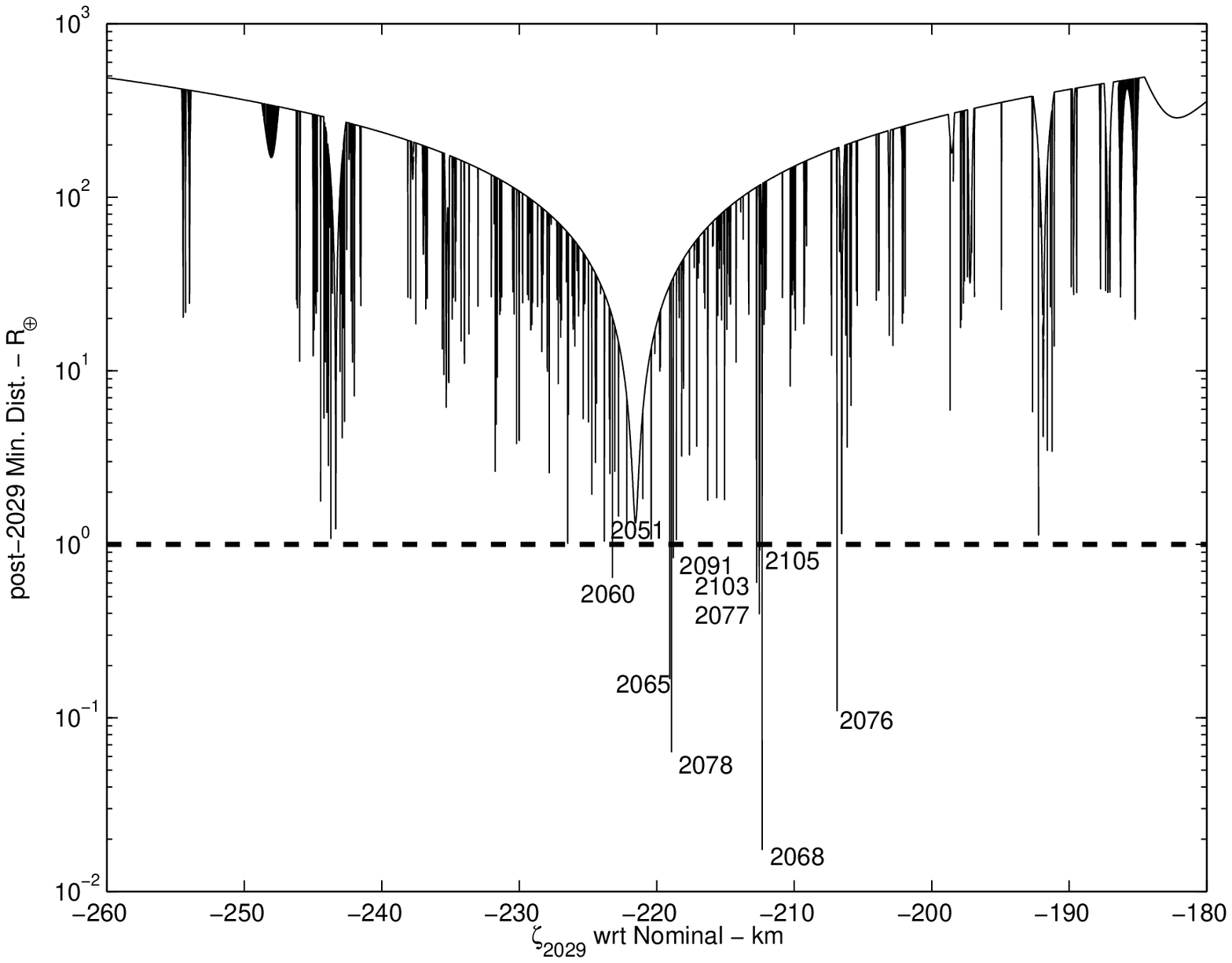}}
\caption{Keyhole map on 2029 $b$-plane. The minimum distance after
  2029 is plotted as a function of $\zeta_{2029}$. The top panel shows
  the full uncertainty region (60000 samples), while the bottom panel is a zoom on the
  2051 impactor complex (200000 samples).}
\label{f:khmap}
\end{figure}

To find the keyholes on the 2029 $b$-plane we propagated a densely
sampled LOV through the year 2113, recording the circumstances of all
close approaches to Earth and in particular the $b$-plane coordinates,
i.e., $(\xi, \zeta)$. On the $b$-plane of a given post-2029 encounter
we can interpolate among nearby samples to identify the minimum
possible encounter distance along the LOV. When this minimum distance
was smaller than 1 Earth radius ($R_\oplus$), we obtained the keyhole
width by mapping back to the 2029 $b$-plane the chord corresponding to
the intersection between the LOV and the Earth cross section. This
technique allows us to identify keyholes in 2029 as small as a few
centimeters in extent (Table~\ref{t:risk}). If we plot the minimum
encounter distance for each LOV sample over the 100 year integration
(and each interpolated local minimum in the encounter distance)
against the associated 2029 $b$-plane coordinate $\zeta_{2029}$, then
we obtain the plot shown in Fig.~\ref{f:khmap}. The upper panel shows
the wider view, which has been extended to the left from nominal to
capture the complex of secondary potential impacts surrounding the
2036 primary resonant return that is situated at about $-1500$ km on
the abscissa. There are other similar collections of potential
impactors in the plot, but the 2036 case is the only one in the
depicted $\zeta$ range where the primary return can lead to impact. As
an example, the lower panel in Fig.~\ref{f:khmap} provides a detailed
view of the complex of potential impactors situated around $-220$ km
on the plot. Here the central post-2029 return, which spawns the
surrounding secondary returns, is in 2051, but no impact is possible
in that year, although an approach closer than $2 R_\oplus$ is
possible.

\begin{figure}[h]
\centerline{\includegraphics[width=14cm]{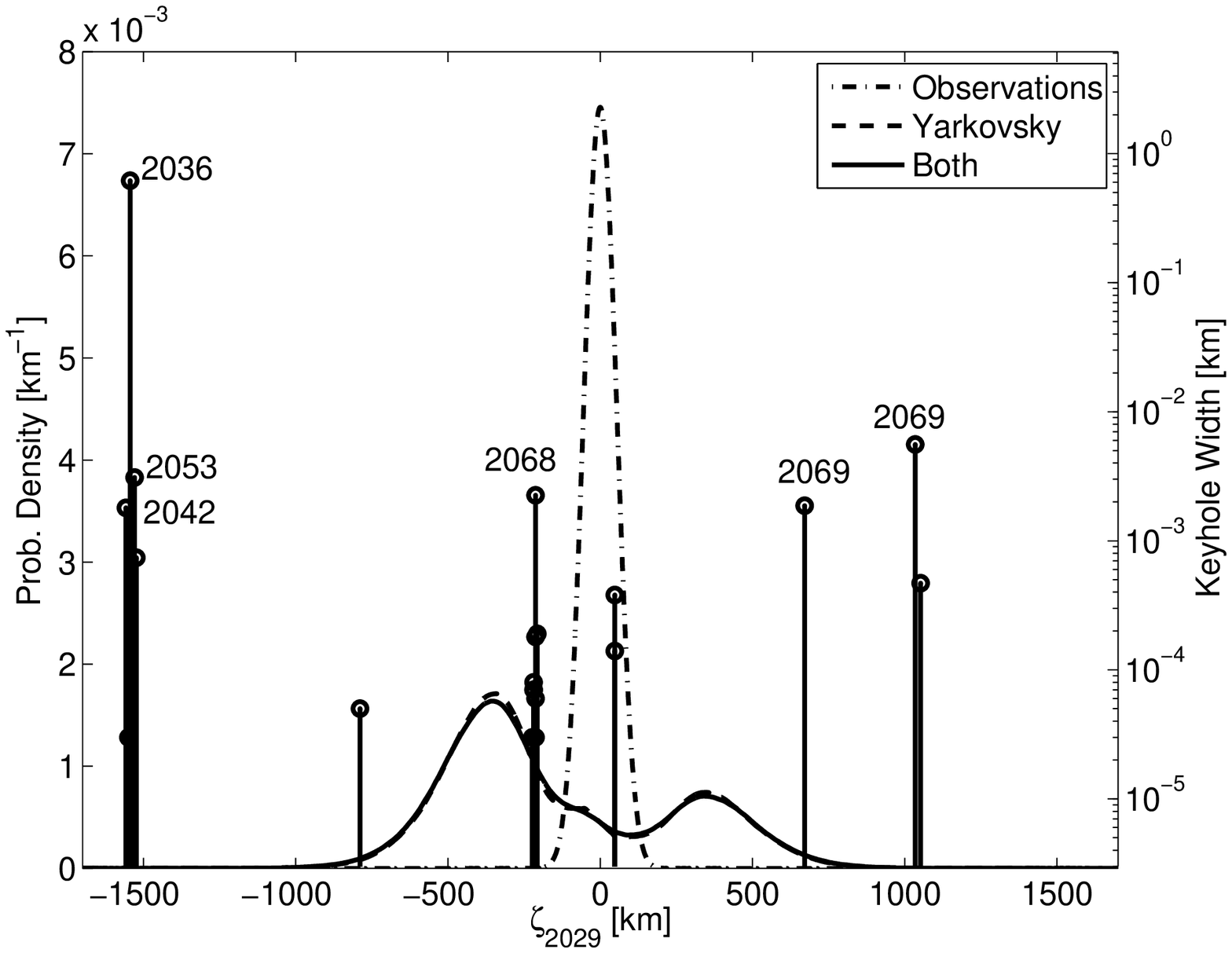}}
\caption{Probability distribution on the 2029 $b$-plane due to the
  uncertainty in the orbital solution, the Yarkovsky effect, and
  both. The origin is set to the nominal $\zeta_{2029} = 47659.24$
  km. Vertical bars correspond to keyholes and the height (in
  logarithmic scale) is proportional to the keyhole width. For
  clarity, only the keyholes with a width larger than 1 m are labeled
  with the year of impact.}
\label{f:zeta}
\end{figure}

Figure~\ref{f:zeta} shows the $\zeta$ distribution on the 2029
$b$-plane by combining 1) the orbital uncertainty obtained from the
orbital fit to the observations neglecting the Yarkovsky effect, and
2) the dispersion due to the Yarkovsky effect. The overall $\zeta$
distribution is therefore obtained as the convolution of the starting
probability density functions. It is interesting to see how the
overall probability density function (PDF) is close to the Yarkovsky
PDF. This is due to the low dispersion of $\zeta$ related to the
orbital uncertainty PDF: if the latter were a Dirac delta function,
the overall PDF would coincide with the Yarkovsky PDF. The figure also
contains the keyholes, where the height of the corresponding bars is
determined by the keyhole width.
The impact probability can be computed as the product of the width of
the keyholes and the PDF. Table~\ref{t:risk} reports the information
on the keyholes along with the associated impact probability.

\begin{table}[t]
\begin{center}
\begin{tabular}{lcccc}
  Date & width & $\Delta\zeta$ wrt TRO & $\sigma$ & IP\\
  TDB & m & km & & \\
  \hline
  2042 Apr. 13.5 & 1.80 & -1557.3 & -4.80 & $1.60\times 10^{-11}$\\
  2041 Apr. 15.3 & 0.03 & -1549.4 & -4.78 & $3.04\times 10^{-13}$\\ 
  2036 Apr. 13.4 & 616.16 & -1543.8 & -4.77 & $7.07\times 10^{-9}$\\ 
  2053 Apr. 12.9 & 3.09 & -1528.9 & -4.73 & $4.91\times 10^{-11}$\\ 
  2044 Apr. 13.3 & 0.74 & -1523.1 & -4.71 & $1.34\times 10^{-11}$\\ 
  2089 Oct. 15.6 & 0.05 & -788.7 & -2.38 & $4.74\times 10^{-9}$\\ 
  2060 Apr. 12.6 & 0.03 & -223.0 & 0.08 & $3.23\times 10^{-8}$\\ 
  2065 Apr. 11.8 & 0.08 & -218.8 & 0.10 & $8.40\times 10^{-8}$\\ 
  2078 Apr. 13.8 & 0.07 & -218.7 & 0.10 & $7.34\times 10^{-8}$\\ 
  2091 Apr. 13.4 & 0.07 & -218.6 & 0.10 & $7.34\times 10^{-8}$\\ 
  {\bf 2103 Apr. 14.4} & {\bf 0.18} & {\bf -212.5} & {\bf 0.12} & $\mathbf{1.82\times 10^{-7}}$\\ 
  2077 Apr. 13.5 & 0.06 & -212.3 & 0.12 & $6.05\times 10^{-8}$\\ 
  2105 Apr. 13.8 & 0.03 & -212.3 & 0.12 & $3.03\times 10^{-8}$\\ 
  {\bf 2068 Apr. 12.6} & {\bf 2.25} & {\bf -212.1} & {\bf 0.12} & $\mathbf{2.27\times 10^{-6}}$\\ 
  {\bf 2076 Apr. 13.0} & {\bf 0.19} & {\bf -206.7} & {\bf 0.13} & $\mathbf{1.85\times 10^{-7}}$\\ 
  2068 Oct. 15.4 & 0.14 &     47.7 &   0.54 & $5.06\times 10^{-8}$\\ 
  {\bf 2105 Oct. 16.3} & {\bf 0.38} &   {\bf   47.7} &  {\bf 0.54} & $\mathbf{1.37\times 10^{-7}}$\\ 
  {\bf 2069 Oct. 16.0} & {\bf 1.87} &   {\bf 671.6} &   {\bf 2.24} & $\mathbf{2.39\times 10^{-7}}$\\ 
  2069 Apr. 13.1 & 5.58 & 1034.5 &   3.53 & $1.39\times 10^{-8}$\\ 
  2077 Apr. 13.6 & 0.47 & 1052.4 &   3.59 & $9.51\times 10^{-10}$\\ 
  \hline
\end{tabular}
\end{center}
\caption{Risk assessment for Apophis. Columns are impact epoch,
  2029 keyhole width and location, inverse of the normal cumulative
  density function, and impact probability. Bold lines correspond to an
  impact probability larger than  Sentry's
  generic completeness, i.e., a value of $8.6 \times 10^{-8}$
  \citep{lov2}.}
\label{t:risk}
\end{table}

The results contained in Table~\ref{t:risk} depend on the assumptions
made on the physical modeling. For example, using a distribution for
which a retrograde rotator is so much more likely than a direct one
could be inaccurate. In general, we expect that the impact
probabilities associated with keyholes close to the peaks of the PDF
should be more stable, while those associated with keyholes in the
tails of the distribution should be more sensitive to changes in the
assumptions because of the exponential decay. To confirm this idea,
Table~\ref{t:ip_sensitivity} reports the values of the impact
probability of the 2036 and 2068 ($\Delta\zeta = -212.14$ km)
keyholes, for different assumptions on Apophis' physical
characterization. We can see that the order of magnitude of
IP$_{2068}$ does not change, while IP$_{2036}$ changes up to three
orders of magnitude.

\begin{table}[t]
\begin{center}
\begin{tabular}{lcccc}
  Assumption & 2036 IP & 2036 $\Delta$IP& 2068 IP & 2068 $\Delta$IP\\
  \hline
  Nominal & $7.07 \times 10^{-9}$ & 0\% & $2.27 \times 10^{-6}$ &
  0\%\\
  Diameter $270 \pm 60$ m & $3.06 \times 10^{-6}$ & +43169\% & $1.68 \times
  10^{-6}$ & -26\%\\
  Unif. $\cos\gamma$ & $2.67 \times 10^{-9}$ & $-62\%$ & $2.14 \times 10^{-6}$ &
  $-6$\%\\
  Symm. $\cos\gamma$ & $1.64 \times 10^{-9}$ & $-77$\% & $1.06 \times 10^{-6}$ &
  $-53$\%\\
  $\sigma_\rho^2$ = 0.6 & $6.60 \times 10^{-7}$ & $+9242$\% & $2.38 \times 10^{-6}$ &
  $+5$\%\\
  $\mu_{d_0}$ = 200 & $1.58 \times 10^{-9}$ & $-78$\% & $3.20 \times 10^{-6}$ &
  $+41$\%\\
  $\mu_{d_0}$ = 400 & $1.59 \times 10^{-8}$ & $+125$\% & $2.02 \times 10^{-6}$ &
  $-11$\%\\
  \hline
\end{tabular}
\end{center}
\caption{Variation of the impact probability for the 2036 and 2068 ($\Delta\zeta = -212.14$ km)
  keyholes for different assumption the Yarkovsky effect modeling:
  nominal assumptions of Sec.\ref{s:yarkovsky}; diameter from
  \citet{delbo_apophis}; uniform distribution for
  $\cos\gamma$; distribution for $\cos\gamma$ symmetric to the one of
  Fig.~\ref{f:physical}; different variance of $\rho$; different means
  of $d_0$
  in the thermal inertia modeling.}
\label{t:ip_sensitivity}
\end{table}

The exceptionally close approach of 2029 may change Apophis' spin
orientation. \citet{scheeres_apophis} show that Apophis could exit the
2029 encounter in complex rotation. Therefore, the post-2029 dynamics
is a function of $\zeta_{2029}$ and post-2029 $A_2$, which we denote
with $A'_2$. In particular, the 2029 keyholes should be considered a
two-dimensional entity in $(\zeta_{2029}, A'_2)$. Then, the IP
of a given keyhole is 
\begin{equation}
IP = \int_{-\infty}^{+\infty} \left( f_1(\zeta_{2029}) w\right) f_2(A'_2) d A'_2
\end{equation}
where $f_1$ and $f_2$ are the PDF of $\zeta_{2029}$ and $A'_2$, and
$w$ is the keyhole width. We checked that for
different values of $A'_2$ the keyhole location changes only slightly
(up to 0.5 km) and the width does not change significantly. Therefore,
IP does not depend significantly on $A'_2$.

\section{Early 2013 apparition}
As already mentioned, the observational dataset available for Apophis
does not allow us to measure the Yarkovsky effect from the orbital
fit. Apophis is scheduled to be observed from the Goldstone radar
observatory in January
2013\footnote{http://echo.jpl.nasa.gov/asteroids/Apophis/Apophis\_2013\_planning.html}
and the Arecibo radar observatory in February
2013\footnote{http://www.naic.edu/$\sim$pradar/sched.shtml}.
Therefore, we wondered whether or not this radar apparition will
constrain the Yarkovsky effect. To answer this question we considered
three different scenarios:
\begin{enumerate}
\item[(a)] no Yarkovsky effect, i.e., $A_2 = 0$ au/d$^2$;
\item[(b)] positive orbital drift with $A_2 = 25 \times 10^{-15}$
  au/d$^2$, which corresponds to the right peak of the $da/dt$
  distribution of Fig.~\ref{f:yarko_par};
\item[(c)] negative orbital drift with $A_2 = -25 \times 10^{-15}$
  au/d$^2$, which corresponds to the left peak of the $da/dt$
  distribution of Fig.~\ref{f:yarko_par}.
\end{enumerate}

\begin{figure}[h]
\centerline{\includegraphics[width=10cm]{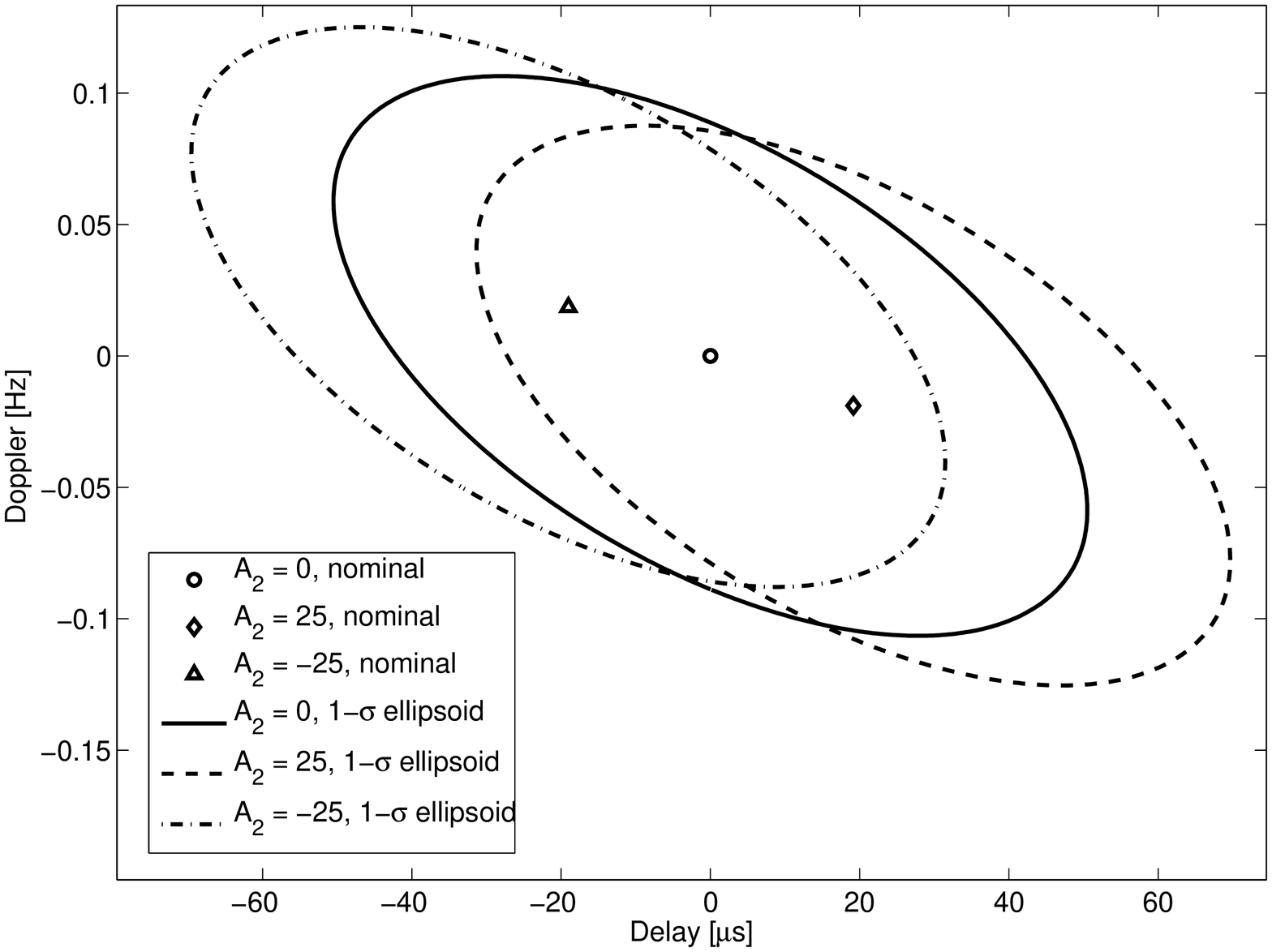}}
\caption{Arecibo delay-Doppler predictions for 2013-02-06.0 for
  different Yarkovsky scenarios. The origin is set to (127.073088 s,
  -56954.4 Hz).}
\label{f:radar_obs}
\end{figure}

Figure~\ref{f:radar_obs} shows the predicted values of delay and
Doppler measurements for 2013 February 6, for the different scenarios
of the Yarkovsky effect. The uncertainty ellipses overlap each other
and their intersection contains all the nominal predictions. We
conclude that the Yarkovsky effect will not be detectable with the
next radar apparition. To further quantify this, we simulated
synthetic delay measurements on 2013 February 5 and 6 (see
Table~\ref{t:radar_obs}). As the relative accuracy of delay
measurements is $\sim$100 times better than the relative accuracy of
Doppler measurements, two delay measurements provide a stronger
orbital constraint than a single delay/Doppler pair. For each scenario
we tried to determine $A_2$ as was done in \citet{yarko_list}. We
found $\sigma_{A_2} = 48 \times 10^{-15}$ au/d$^2$ with the simulated
radar data, which is about two times larger than a reasonable value of
$|A_2|$.

This result may look surprising, since in the cases of (6489) Golevka
\citep{golevka} and 1999 RQ$_{36}$ \citep{chesley_rq36} the
availability of three radar apparitions allowed strong detections of
the Yarkovsky effect. However, in these cases the time span covered by
radar observations was 12 yr (1991 to 2003 for Golevka, 1999 to 2011
for 1999 RQ$_{36}$) while for Apophis it is 8 yr. Moreover, the second
radar apparitions of Golevka in 1995 and 1999 RQ$_{36}$ in 2005 were
approximately in the middle of the other two, while the second
apparition of Apophis was in 2006, just one year after the first
apparition, and contains only one Doppler measurement, which is not as
precise as delay measurements. Therefore, both the timing and
precision of the data may explain the failure in detecting the
Yarkovsky effect with three radar apparitions. As a matter of fact,
the 2005, 2006, and 2013 datasets essentially contain the same
information as two radar apparitions, since only 2005 and 2013 contain
delay measurements.

\begin{table}[t]
\begin{center}
\begin{tabular}{cccc}
  Scenario & $A_2$ & Date & Delay\\
  & $10^{-15}$ au/d$^2$ & UTC & s\\
  \hline
  (a) &        0 & 2013-02-05.0 & 124.886779(1)\\
  (a) &        0 & 2013-02-06.0 & 127.073088(1)\\
  (b) &   25.0 & 2013-02-05.0 & 124.886797(1)\\
  (b) &   25.0 & 2013-02-06.0 & 127.073107(1)\\
  (c) & -25.0 & 2013-02-05.0 & 124.886760(1)\\
  (c) & -25.0 & 2013-02-06.0 & 127.073069(1)\\
  \hline
\end{tabular}
\end{center}
\caption{Simulated February 2013 delay observations from Arecibo for
  different values of $A_2$. The uncertainty is assumed to be 1
  $\mu$s.}
\label{t:radar_obs}
\end{table}

At this point it is interesting to see what would be the effect of the
January/February 2013 radar apparition on the risk assessment. As the
Yarkovsky effect will not be directly detected, the main contribution
will be on the orbital uncertainty information. For example, let us
consider scenario (c), i.e., we used the radar observations simulated
according to the assumed $A_2 = -25 \times 10^{-15}$ au/d$^2$.  We
computed the corresponding orbital solution without accounting for the
Yarkovsky effect, which is not measurable. The corresponding 2029
$b$-plane coordinates are $(\xi,\zeta) = (9485.10,47369.09)$ km, with
uncertainties 5.30 km and 11.00 km, respectively. While the
uncertainty in $\xi$ is almost the same, the uncertainty in $\zeta$
decreased by a factor of 4 and then the corresponding PDF of $\zeta$
due to the orbital uncertainty is even closer to a Dirac delta
function. As a consequence, the Yarkovsky PDF and the overall PDF have
almost the same curve. The changes in the IP are mostly due to a shift
of about -300 km of the nominal Yarkovsky-free prediction. All the
virtual impactors \citep{lov2} of Table~\ref{t:risk} persist, and four
of them now have an IP$>8.6 \times 10^{-8}$: IP$_{2036} = 2.23 \times
10^{-7}$, IP$_{2068}(\Delta\zeta=-212.14 km) = 6.95 \times 10^{-7}$,
IP$_{2068}(\Delta\zeta=47.68 km) = 1.03 \times 10^{-7}$, and
IP$_{2105} = 2.79 \times 10^{-7}$.


Though the Yarkovsky effect will not be constrained, radar and optical
photometry can provide independent information to improve Apophis'
physical modeling. In particular, it may be possible to further
constrain the diameter and the albedo, and to estimate the spin
orientation, which would allow us to remove the bimodality in the
Yarkovsky distribution.

The next opportunity for radar observations is the September 2021
close approach. Meanwhile, ground based optical observations are
possible and can help in determining the Yarkovsky effect before the
September 2021 close approach.  Table~\ref{t:optical_yarko} contains
the uncertainty in the $A_2$ determination by assuming simulated radar
observations in February 2013 and progressively adding simulated
ground based optical observations weighted at 0.15". We conclude that
the Yarkovsky effect would be significantly constrained between late
2020 and early 2021.

\begin{table}[t]
\begin{center}
\begin{tabular}{cccc}
  Dates & $\sigma_{A_2}$ [$10^{-15}$ au/d$^2$]\\
  \hline
  Jan. 2013 -- Jun. 2013 & 47\\
  Sep. 2013 -- May 2015 & 40\\
  Oct. 2015 -- Jan. 2016 & 39\\
  Mar. 2018 -- May 2018 & 32\\
  Dec. 2018 -- Aug. 2020 & 16\\
  Nov. 2020 -- May 2021 & 4\\
  \hline
\end{tabular}
\end{center}
\caption{Uncertainty in the Yarkovsky effect determination as a
  function of increasing number of simulated optical observations
  weighted at 0.15". For each interval we simulated one observation
  every two months.}
\label{t:optical_yarko}
\end{table}

\section{Conclusions}
In this paper we presented an updated and careful impact risk analysis
for asteroid (99942) Apophis. The presence of the scattering close
approach in 2029 calls for accurate orbit estimation and dynamical
modeling. Therefore, we selected astrometry for which the uncertainty
information is rigorously computed by the observer and added the
contributions of observatories with comparable performance.  To model
the Yarkovsky effect we performed a Monte Carlo simulation relying on
the best physical model currently available.

We obtained the distribution on the 2029 $b$-plane by combining the
orbital uncertainty and the dispersion due to the Yarkovsky effect. We
then found twenty keyholes corresponding to future impacts and
computed the corresponding impact probabilities. In particular, we
found a $2.27 \times 10^{-6}$ impact probability for an impact in
2068.

We analyzed the stability of the computed impact probabilities with
respect to the assumed Apophis' physical characterization. Impacts
corresponding to keyholes close to the nominal solution have a stable
impact probability (the order of magnitude does not change). On
the other hand, for keyholes in the tails of the distribution we found
variations up to three orders of magnitude.

We also showed that the February 2013 radar apparition will not allow
a direct detection of the Yarkovsky effect, and therefore the
post-2029 possibility of impact will not be ruled out. The main
contribution will be a reduction of the orbital uncertainty, thus
making the dispersion due to the Yarkovsky the predominant source of
uncertainty.

\section*{Acknowledgments}
DF was supported for this research by an appointment to the NASA
Postdoctoral Program at the Jet Propulsion Laboratory, California
Institute of Technology, administered by Oak Ridge Associated
Universities through a contract with NASA.

SC and PC conducted this research at the Jet Propulsion Laboratory,
California Institute of Technology, under a contract with NASA.

Copyright 2013. All rights reserved.
\bibliography{apophis}
\bibliographystyle{elsarticle-harv}
\end{document}